\documentclass[12pt,thmsa]{article}
\usepackage{amssymb}

\usepackage{sw20lart}

\input{tcilatex}
\begin{document}

\author{Emilio Santos \and Departamento de F\'{i}sica, Universidad de Cantabria,
\and 39005 Santander, Spain}
\title{Bell inequalities for two-photon experiments testable at low detection
efficiency without assuming fair sampling }
\date{December, 27, 2006}
\maketitle

\begin{abstract}
A family of local models containing two angles as hidden variables is
defined for experiments measuring polarization correlation of optical
photons. Searching for the best model of the family, that is giving
predictions most close to quantum mechanics, allows deriving Bell-type
inequalities which may be tested with relatively low detection efficiency.

\textbf{PACS. }03.65.Ud Entanglement and quantum nonlocality (e.g. EPR
paradox, Bell\'{}s inequalities, GHZ states, etc.) - 42.50.Ar Photon
statistics and coherence theory
\end{abstract}

\section{\protect\smallskip Introduction}

More than forty years have elapsed since John Bell\cite{Bell} proposed his
celebrated inequalities. These inequalities, which involve measurable
quantities, provide necessary conditions for local realism and, in some
experiments with ideal set-ups, contradict the predictions of quantum
mechanics. Many empirical tests have been performed of local realism against
quantum mechanics, via the Bell inequalities, but no experiment has been
conclusive. In fact although the results have generally agreed with the
predictions of quantum mechanics, no experiment has given results
incompatible with local realism, as is shown by the existence of local
realistic models for all of them\cite{SantosFP}. The inability to perform a
true empirical test of local realism is commonly disguised with the claim
that it has already been refuted by the experiments, modulo some irrelevant
loopholes. But I think that the extreme difficulty to make a loophole-free
test, proved by the unsuccessful effort of forty years, does not support the
common wisdom that the question of local realism is settled. On the contrary
the conclusion is that further research is needed.

Locality should be understood in the relativistic sense, that is as
impossibility of superluminal communication. Thus testing locality requieres
measurements made in regions \textit{spacially }separated, in the sense of
relativity theory. As a consequence the tests are extremely difficult with
massive particles\cite{Santos95} and reliable experiments must be performed
with photons. On the other hand, tests with high energy photons are not
possible due to the lack of efficient polarization analyzers. These
difficulties have caused that most of the experimental tests have been
performed with optical photons, where good polarization analyzers exist and
locality may be insured. However these experiments suffer from a detection
loophole, due to the fact that a good overall detection efficiency has not
yet been achieved. In fact, it is well known that efficiencies as high as
80\% are required for loophole-free tests of local realism. Detectors with
high quantum efficiency already exist, but there are other difficulties
reducing the overall efficiency to about 30\% or less in practice.

It is common to formulate the question of local realism in terms of local
hidden-variables (LHV) theories. That is, an experiment refutes local
realism if there is no LHV model compatible with their results. Bell
inequalities are necessary conditions for LHV theories but, as said above,
they are very difficult to test. Actually all inequalities empirically
violated till now are not genuine Bell inequalities, derived from the
conditions of realism and locality alone, but inequalities whose derivation
requires auxiliary assumptions. It is rather obvious that violations of such
inequalities do not refute the whole family of LHV theories but only
restricted families, namely those fulfilling the auxiliary assumptions. In
my opinion the families of local hidden variables theories so far refuted by
the experiments are rather unplausible\cite{SantosFP}. This is the case, in
particular, for those fulfilling the fashionable ``fair sampling
hypothesis''. This is the assumption that the photons actually detected are
representative of the whole set of photons emitted. But the role of hidden
variables is precisely to distinguish, from each other, several physical
systems in the same \textit{pure} quantum state, that is systems which are
identical according to the standard interpretation of quantum mechanics.
Thus in any LHV model of an experiment it is natural to assume that the
photons detected and those not detected correspond to different values of
the hidden variables and consequenly the sample of detected photons is not
representative of the whole set. In conclusion, the fair sampling assumption
amounts at dismissing all sensible hidden variables theories from the start.

In the present paper I study a new family of LHV models which is rather
natural, in my opinion, and it allows the derivation of inequalities able to
discriminate between the said family and quantum mechanics. The inequalities
are easily testable in optical experiments, in particular the tests would
require only moderate detection efficiencies, of the order of 30\%. The said
family was already considered in a previous publication\cite{Santos}, where
an inequality fulfilled by some sub-family was proposed. That inequality has
been tested empirically with the result that it was fulfilled and the
quantum predictions contradicted although, according to the authors of the
experiment\cite{Genovese}, the contradiction cannot be considered a
violation of quantum mechanics. In the present paper I propose inequalities
which should hold true for \textit{all members} of the said LHV family, but
are contradicted by the quantum predictions in some cases.

\section{A natural family of local hidden variables models}

For the sake of clarity I shall consider experiments measuring the
polarization correlation of optical photon pairs, although the
generalization to other cases is possible. The set-up consists of a source
of photon pairs each member of the pair travelling in a different direction,
crossing a lens system, a polarization analyzer and arriving at a detector.
If the polarization planes of the analyzers are determined by the angles $%
\phi _{1\text{ }}$and $\phi _{2\text{ }},$ respectively, the results of the
experiment may be summarized in two single rates, $R_{1}(\phi _{1\text{ }})$
and $R_{2}(\phi _{2\text{ }})$ , and a coincidence rate $R_{12}(\phi _{1%
\text{ }},\phi _{2\text{ }}).$ (In recent experiments four coincidence
rates, rather than one, are measured because two-channel polarizers are
used, a situation which will be considered below). In a polarization
correlation experiment the detection rates should be obtained from
appropriately defined probabilities $p_{1},p_{2\text{ }}$and $p_{12}$, that
is 
\begin{equation}
R_{j}(\phi _{j\text{ }})=R_{0}p_{j}(\phi _{j\text{ }}),\;j=1,2,\;R_{12}(\phi
_{1\text{ }},\phi _{2\text{ }})=R_{0}p_{12}(\phi _{1\text{ }},\phi _{2\text{ 
}}),  \label{02}
\end{equation}
where $R_{0}$ is the production rate of photon pairs in the source, a
quantity not measurable in standard experiments. (Here it is assumed that
all photocounts come from photons produced in pairs in the source. In
practice there may be counts of a different origin, e.g. dark counts in each
detector, but they will be neglected for the moment, see section 4 below).
Following Bell a LHV model consists of three functions, $f(\lambda
),Q_{1}(\lambda ,\phi _{1}),Q_{2}(\lambda ,\phi _{2}),$ where $\lambda $
stands for one or several hidden variables, such that the detection
probabilities could be obtained by means of the integrals 
\begin{equation}
p_{j}(\phi _{j\text{ }})=\int f(\lambda )Q_{j}(\lambda ,\phi _{j})d\lambda
,\;p_{12}(\phi _{1\text{ }},\phi _{2\text{ }})=\int f(\lambda )Q_{1}(\lambda
,\phi _{1})Q_{2}(\lambda ,\phi _{2})d\lambda .\;  \label{03}
\end{equation}
The essential requirements of realism and locality imply that the said
functions fulfil the conditions 
\begin{equation}
f(\lambda )\geq 0,\int f(\lambda )d\lambda =1,\;0\leq Q_{j}(\lambda ,\phi
_{j})\leq 1.  \label{04}
\end{equation}

A natural, but relatively simple, family of local hidden variables model is
obtained if we assume that the set $\lambda $ of hidden variables may be
written $\lambda \equiv \left\{ \chi _{1},\mu _{1},\chi _{2},\mu
_{2}\right\} ,$ where $\chi _{1}$ and $\mu _{1}$ ($\chi _{2}$ and $\mu _{2})$
are variables of the first (second) photon of a pair and $\chi _{j}$ is a
polarization angle, so that $\chi _{j}$ and $\chi _{j}+\pi $ represent the
same polarization. Actually we may assume that $\mu _{1}$ and $\mu _{2}$
label a set of variables each, rather than a single one. In principle the
function $f(.)$ gives the correlation amongst the four (sets of) variables.
But simplifies a lot the model to assume that the variables, or sets of
variables, $\mu _{1}$ and $\mu _{2}$ are uncorrelated amongst themselves
(this is a consequence of locality) and uncorrelated with the polarization
angles (this is an assumption of simplicity), so that the functions $%
f(\lambda )$ and $Q_{j}(\lambda ,\phi _{j})$ may be written 
\begin{equation}
f(\lambda )\equiv \rho (\chi _{1},\chi _{2})g_{1}\left( \mu _{1}\right)
g_{2}\left( \mu _{2}\right) ,Q_{j}(\lambda ,\phi _{j})\equiv Q_{j}(\chi
_{j},\mu _{j},\phi _{j}),  \label{05}
\end{equation}
where $\rho $ and $g_{j}$ are positive and normalized. When eqs.$\left( \ref
{05}\right) $ are inserted in $\left( \ref{03}\right) $ and the integrals in 
$\mu _{j}$ performed we get 
\begin{eqnarray}
p_{j}(\phi _{j\text{ }}) &=&\int \rho (\chi _{1},\chi _{2})P_{j}(\chi
_{j},\phi _{j})d\chi _{1}d\chi _{2},\;  \nonumber \\
p_{12}(\phi _{1\text{ }},\phi _{2\text{ }}) &=&\int \rho (\chi _{1},\chi
_{2})P_{1}(\chi _{1},\phi _{1})P_{2}(\chi _{2},\phi _{2})d\chi _{1}d\chi
_{2},\;  \label{06}
\end{eqnarray}
where 
\begin{equation}
P_{j}(\chi _{j},\phi _{j})=\int g_{j}(\mu _{j})Q_{j}(\chi _{j},\mu _{j},\phi
_{j})d\mu _{j}.  \label{07}
\end{equation}
Thus our model requires defining only the functions $\rho $ and $P_{j}$,
which fulfil conditions of positivity and normalization similar to those of $%
f$ and $Q_{j},$ respectively, in eqs.$\left( \ref{04}\right) .$

In order to derive testable inequalities from our LHV model we shall
consider firstly experiments where there is rotational symmetry, which is
common at least approximately in actual experiments. That is, we assume that
the single rates $R_{j}$ do not depend on the orientation of the polarizers
(i. e. the angle $\phi _{j})$ and the coincidence rate depends only on the
difference $\phi _{1}-\phi _{2}=\phi .$ We shall also assume symmetry
amongst the polarizer-detector systems so that $P_{1}=P_{2}$. (See section 4
below for the study of cases where these symmetries do not hold true). In
these conditions it is appropriate to replace eqs.$\left( \ref{06}\right) $
by 
\begin{eqnarray}
p_{12}(\phi ) &=&\int \rho (\chi _{1}-\chi _{2})P(\chi _{1}-\phi _{1})P(\chi
_{2}-\phi _{2})d\chi _{1}d\chi _{2},  \label{1} \\
p_{j} &=&\int \rho (\chi _{1}-\chi _{2})P(\chi _{j}-\phi _{j})d\chi
_{1}d\chi _{2},\;j=1,2.\;  \label{1a}
\end{eqnarray}
where here and below all functions are periodic with period $\pi $ and the
integrals go from -$\pi /2$ to $\pi /2.$ In addition the functions $\rho $
and $P$ possess the following properties of positivity, symmetry and
normalization ($\rho $ is normalized so that $p_{12}(\phi )=1$ if $P=1$ in $%
\left( \ref{1}\right) )$%
\begin{equation}
\rho (x)=\rho (-x)\geq 0,\int \rho (x)dx=1/\pi ,\;0\leq P(x)=P(-x)\leq 1.
\label{2}
\end{equation}
I shall add the following two conditions which are plausible on physical
grounds 
\begin{equation}
\;\frac{d\rho \left( x\right) }{d\left| x\right| }\leq 0,\;\frac{dP\left(
x\right) }{d\left| x\right| }\leq 0.  \label{2a}
\end{equation}
The first inequality means that the pairs where $\chi _{1}=\chi _{2}$ are
most likely produced in the source (this equalitity may correspond to
parallel or perpendicular polarization, depending on the actual experiment).
The second inequality means that the detection is most probable when the
incoming photon has the polarization close to the plane of the analyzer. Eqs.%
$\left( \ref{1}\right) $ to $\left( \ref{2a}\right) $ define our ``natural''
family of LHV models.

For later convenience I make the change of variables 
\begin{equation}
\chi _{1}-\phi _{1}=u,\;\chi _{2}-\phi _{2}=v,\;\phi _{1}-\phi _{2}=\phi ,
\label{3}
\end{equation}
which leads to 
\begin{equation}
p_{12}(\phi )=\int \rho (u-v+\phi )P(u)P(v)dudv.  \label{4}
\end{equation}
Hence we get the following results 
\begin{equation}
\int p_{12}(\phi )d\phi =p_{1}p_{2},\;p_{1}=p_{2}=C_{0}/\pi ,\;\int
p_{12}(\phi )\cos 2\phi \,\,d\phi \equiv C_{1}^{2}/\pi ,  \label{9}
\end{equation}
where the constants $C_{k}$ are defined by 
\begin{equation}
C_{k}\equiv \int P(x)\cos \left( 2kx\right) dx.  \label{9a}
\end{equation}
It is also convenient to introduce the new function 
\begin{equation}
f(y)=f(-y)=\int P(x+\frac{y}{2})P(x-\frac{y}{2})dx.  \label{5}
\end{equation}
whence we get 
\begin{equation}
p_{12}(\phi )=\int \rho (y+\phi )f(y)dy=\int \rho (y)f(y-\phi )dy,  \label{6}
\end{equation}
the latter equality following from the periodicity of the functions involved.

\section{Comparison with the quantum predictions}

Now I will investigate whether the quantum predictions are compatible with
the family of LHV models above defined. Quantum mechanics predicts, for
experiments with rotational symmetry and both detectors having the same
efficiency, $\eta ,$ 
\begin{equation}
p_{j}^{Q}=\frac{1}{2}\eta ,\;p_{12}^{Q}\left( \phi \right) =\frac{1}{4}\eta
^{2}\left( 1+V\cos 2\phi \right) .  \label{16}
\end{equation}
Our family of models $\left( \ref{1}\right) $ agrees with the
quantum-mechanical prediction for the single probabilities, $p_{j},$
provided we choose 
\begin{equation}
C_{0}=\int P(x)dx=\frac{\pi \eta }{2}.  \label{14b}
\end{equation}
Then I will search for the best LHV model of the form $\left( \ref{1}\right) 
$, defining ``best'' by the condition that, for fixed $\eta $ and $V,$ the
prediction for the coincidence probability is as close as possible to the
quantum prediction. That is the quantity S must be a minimum, where 
\begin{equation}
S\equiv \int d\phi \left[ p_{12}(\phi )-\frac{1}{4}\eta ^{2}\left( 1+V\cos
2\phi \right) \right] ^{2},  \label{7}
\end{equation}
with $p_{12}$($\phi )$ given by $\left( \ref{6}\right) $ and $\left( \ref{5}%
\right) $ and the functions $\rho (x)$ and $P(x)$ fulfilling the conditions $%
\left( \ref{2}\right) .$ For the solution of the problem I shall proceed in
two steps. Firstly$,$ for a given $P(x)$ fulfilling $\left( \ref{14b}\right)
,$ I search for the \textit{best }positive and normalized function $\rho (x)$%
. In the second step I shall look for the \textit{best} $P(x).$

In the first step\textit{\ }fixing $P(x)$ amounts at fixing the function $%
f(y)$ (see eq.$\left( \ref{5}\right) $). For the solution of the variational
problem, eq.$\left( \ref{7}\right) ,$ the positivity of $\rho $ is insured
writing 
\[
\rho (x)=\psi (x)^{2},\psi \in \mathbf{R,}
\]
and the normalization constraint may be taken into account by means of a
Lagrange parameter $\lambda $ . Thus the problem becomes 
\[
\delta \left\{ S\left[ \psi (x)^{2}\right] +\lambda \int \psi
(x)^{2}dx\right\} =0.
\]
Hence we get, after some algebra, that either $\psi (x)=0$ or 
\begin{equation}
\int \psi (x)^{2}dx\int d\phi f(x-\phi )f(y-\phi )=\lambda +\frac{1}{4}\eta
^{2}\int d\phi f(y-\phi )[1+V\cos 2\phi ].  \label{12}
\end{equation}
For the solution of this integral equation the properties of $\rho (x),$ see 
$\left( \ref{1}\right) ,$ suggest the Fourier expansion\textit{\ } 
\begin{equation}
\psi (x)^{2}=\sum_{k=0}^{\infty }A_{k}\cos (2kx),  \label{12b}
\end{equation}
which, inserted in eq.$\left( \ref{12}\right) $ leads to (see eq.$\left( \ref
{9a}\right) )$%
\begin{equation}
\sum_{k=0}^{\infty }A_{k}C_{k}^{2}\cos (2ky)=\lambda +\frac{1}{4}\eta
^{2}\left[ C_{0}^{2}+VC_{1}^{2}\cos (2y)\right] .  \label{12d}
\end{equation}
Here I have used the following relation, easily derivable from the
properties of the function $f(x)$ (see eq.$\left( \ref{5}\right) ),$ 
\[
\int f(u-v)\cos (2ku)du=C_{k}^{2}\cos (2kv).
\]
Eq.$\left( \ref{12d}\right) $ holds true for any $y$ if 
\begin{equation}
A_{0}=\frac{4\lambda +\eta ^{2}C_{0}^{2}}{4C_{0}^{4}},\;A_{1}=\frac{\eta
^{2}V}{4C_{1}^{2}},\;A_{k}=0\text{ for }k\geq 2.  \label{12a}
\end{equation}
Thus the function $\rho \left( x\right) $ will be 
\begin{equation}
\rho (x)=\frac{4\lambda +\eta ^{2}C_{0}^{2}}{4C_{0}^{4}}\left[ 1+\frac{\eta
^{2}C_{0}^{4}V}{C_{1}^{2}\left( 4\lambda +\eta ^{2}C_{0}^{2}\right) }\cos
2x\right] _{+},  \label{13}
\end{equation}
where $[.]_{_{+}}$ means putting 0 if the quantity inside the parenthesis is
negative and $\lambda $ should be calculated from the normalization
condition (see eq.$\left( \ref{2}\right) )$. There are two cases which must
be analyzed separately.

The first case corresponds to the function $P(x)$ fulfilling 
\begin{equation}
\frac{C_{1}}{C_{0}}=\frac{\int P(x)\cos 2x\,dx}{\int P(x)\,dx}\geq \sqrt{V}.
\label{14}
\end{equation}
Then we may take $\lambda =0$ leading to 
\begin{equation}
\rho (x)=\frac{1}{\pi ^{2}}\left[ 1+\frac{C_{0}^{2}}{C_{1}^{2}}V\cos
2x\right] =\frac{1}{\pi ^{2}}\left[ 1+\frac{\pi ^{2}\eta ^{2}}{4C_{1}^{2}}%
V\cos 2x\right] ,  \label{14a}
\end{equation}
where the choice of $C_{0}$, eq.$\left( \ref{14b}\right) ,$ has been taken
into account. This gives perfect agreement with the quantum predictions, $%
\left( \ref{16}\right) ,$ for both single and coincidence probabilities.

It is interesting to know the range of values of $\eta $ and $V$ where it is
possible the agreement between the model and quantum predictions. Actually
the constraints $\left( \ref{2}\right) $ and $\left( \ref{14b}\right) $ put
an upper bound to the left hand side of the inequality $\left( \ref{14}%
\right) .$ It is not difficult to realize that the best function $P(x)$,
that is the one saturating the bound, is 
\begin{equation}
P(x)=1\text{ if }\left| x\right| \leq \pi \eta /4,\text{ }0\text{ otherwise.}
\label{21}
\end{equation}
This choice leads to 
\[
C_{1}=\sin \left( \pi \eta /2\right) , 
\]
whence $\left( \ref{14}\right) $ becomes 
\begin{equation}
V\leq V_{\max }=\frac{\sin ^{2}\left( \pi \eta /2\right) }{\left( \pi \eta
/2\right) ^{2}}\simeq 1-\frac{\pi ^{2}\eta ^{2}}{12},  \label{20}
\end{equation}
the latter approximation being valid for $\eta <<1.$ We stress that, if this
inequality is fulfilled, there are LHV models of the type $\left( \ref{1}%
\right) $ giving complete agreement with quantum mechanics. In fact, the
choice eq.$\left( \ref{14a}\right) $ for $\rho \left( x\right) $ and eq.$%
\left( \ref{21}\right) $ for $P(x)$ gives the desired result.

We pass to analyze the second of the two cases in eq.$\left( \ref{13}\right) 
$ which, by the arguments of the previous paragraph, will correspond to the
violation of the inequality $\left( \ref{20}\right) .$ In this case the
function $\rho (z)$ may be written, correctly normalized according to eq.$%
\left( \ref{2}\right) ,$ 
\begin{equation}
\rho (z)=\frac{1}{\pi }\left[ \pi +\tan (2\varepsilon )-2\varepsilon \right]
^{-1}\left( 1+\frac{\cos 2z}{\cos 2\varepsilon }\right) _{+},\varepsilon \in
\left( 0,\frac{\pi }{4}\right)  \label{15}
\end{equation}
and I will search for the departures between the model and
quantum-mechanical predictions using that function. For the single detection
probability, $p_{j},$ there is no departure, that is the predictions exactly
agree, provided we make the choice $\left( \ref{14b}\right) $. However there
is disagreement for the coincidence probability, which may be most
conveniently exhibited expanding $\left( \ref{15}\right) $ in a Fourier
series of the form 
\begin{equation}
\rho \left( z\right) =\frac{1}{\pi ^{2}}\sum_{n=0}^{\infty }a_{n}\cos \left(
2nz\right) .  \label{15a}
\end{equation}
We get $a_{0}=1,$ as it should $\rho \left( z\right) $ being normalized in
the sense of $\left( \ref{2}\right) $, and 
\begin{eqnarray}
a_{1} &=&\left[ \pi +\tan (2\varepsilon )-2\varepsilon \right] ^{-1}\left[ 
\frac{\pi -2\varepsilon }{\cos \left( 2\varepsilon \right) }+\sin \left(
2\varepsilon \right) \right] =1+2\varepsilon ^{2}-\frac{8\varepsilon ^{3}}{%
\pi }+O(\varepsilon ^{4}),  \nonumber \\
a_{n} &=&\frac{2}{n(n^{2}-1)}\left( -1\right) ^{n}\frac{\sin \left(
2n\varepsilon \right) -n\tan \left( 2\varepsilon \right) \cos \left(
2n\varepsilon \right) }{\pi +\tan (2\varepsilon )-2\varepsilon }  \label{22}
\\
&=&\left( -1\right) ^{n}\frac{16}{3\pi }\varepsilon ^{3}+O(\varepsilon ^{5})%
\text{ for n}\geq 2,  \nonumber
\end{eqnarray}
Hence it is straightforward to get the model predictions for $p_{12}$ from $%
\left( \ref{4}\right) $ and $\left( \ref{21}\right) $, that is 
\begin{equation}
p_{12}\left( \phi \right) =\frac{1}{4}\eta ^{2}\left[ 1+\sum_{n=1}^{\infty
}a_{n}\frac{C_{n}^{2}}{\left( \pi \eta /2\right) ^{2}}\cos (2n\phi )\right] ,
\label{22a}
\end{equation}
where $C_{n}$ was defined in eq.$\left( \ref{9a}\right) .$

Our aim is the find the best local model, in the sense of S $\left( \ref{7}%
\right) $ being a minimum for given $\eta $ with $P(x)$ and $\rho (z)$
fulfilling, respectively, eqs.$\left( \ref{14b}\right) $ and $\left( \ref{15}%
\right) .$ It is not difficult to realize that the minimum S will correspond
to $p_{12}(\phi )$ being of the form 
\begin{equation}
p_{12}\left( \phi \right) =\frac{1}{4}\eta ^{2}[1+V\cos \left( 2\phi \right)
+\delta \left( \phi \right) ],  \label{23}
\end{equation}
with $\delta \left( \phi \right) $ containing only terms with $n\geq 2$ in
the Fourier expansion $\left( \ref{22a}\right) $. For any choice of $P(x)$
this will be the case if the following relation between $\varepsilon ,\eta
,V $ and $C_{1}$ holds true 
\begin{equation}
V=\frac{C_{1}^{2}}{\left( \pi \eta /2\right) ^{2}}a_{1}.  \label{24}
\end{equation}

Now we shall choose $P(x)$ so that $\left| \delta \left( \phi \right)
\right| $ is as small as possible. From eqs.$\left( \ref{22}\right) $ we see
that this requires that $\varepsilon $ is small which, as $a_{1}$ increases
with $\varepsilon $, implies that $C_{1}$ must be high (see eq.$\left( \ref
{24}\right) )$. Hence the best choice for $P(x)$ will correspond to the
maximum possible value of $C_{1}$ compatible with eq.$\left( \ref{14b}%
\right) ,$which happens if $P(x)$ is chosen as in eq.$\left( \ref{21}\right) 
$. After that the value of $\varepsilon $ may be obtained from eq.$\left( 
\ref{24}\right) $ giving, to order $O(\varepsilon ^{2}),$ 
\begin{equation}
V=(1+2\varepsilon ^{2})\frac{\sin ^{2}\left( \pi \eta /2\right) }{\left( \pi
\eta /2\right) ^{2}}+O(\varepsilon ^{3})\Longrightarrow \varepsilon \simeq 
\sqrt{\frac{1}{2}\left( V-\frac{\sin ^{2}\left( \pi \eta /2\right) }{\left(
\pi \eta /2\right) ^{2}}\right) _{+},}  \label{24a}
\end{equation}
an equality clearly showing that $\varepsilon $ is a measure of the
violation of the inequality $\left( \ref{20}\right) $ ($\varepsilon =0$ if
the inequality holds true).

With the said choices of $P(x)$ and $\rho \left( x\right) $ the departure of
the model from the quantum predictions is given by 
\begin{equation}
\delta \left( \phi \right) =\sum_{n=2}^{\infty }a_{n}\frac{\sin ^{2}\left(
n\pi \eta /2\right) }{\left( n\pi \eta /2\right) ^{2}}\cos \left( 2n\phi
\right) .  \label{28}
\end{equation}
Inserting here the expressions for $a_{n}$, eq.$\left( \ref{22}\right) ,$ it
is straightforward, although lengthy, to get the expression of $\delta
\left( \phi \right) .$ However the parameter $\varepsilon $ will be very
small in practice (i. e. $\varepsilon <<\pi /4),$ which allows working to
lowest order in $\varepsilon ,$ that is $O(\varepsilon ^{3})$ (see $\left( 
\ref{22}\right) ).$ Although that approximation of $a_{n}$ is good only for
the terms with small $n$, the terms with high $n$ contribute but slightly.
We may get the explicit form of $\delta \left( \phi \right) $ to order $%
O(\varepsilon ^{3})$ from eq.$\left( \ref{28}\right) $ with $a_{n}$ given by
eq.$\left( \ref{22}\right) $. We shall use the summation formula 
\[
\sum_{n=1}^{\infty }\frac{\left( -1\right) ^{n}}{n^{2}}\cos \left( nx\right)
=\frac{x^{2}}{4}-\frac{\pi ^{2}}{12},\text{ }x\in \left[ -\pi ,\pi \right] , 
\]
but we must substitute $\left( -1\right) ^{n}cos(n(x-\pi ))$ for $cos(nx)$
whenever $x>\pi $ (but $x\leq 2\pi )$ and a similar change if $x<-\pi .$ In
this case the summation formula needed is 
\[
\sum_{n=1}^{\infty }\frac{1}{n^{2}}\cos \left( nx\right) =\frac{\left( x-\pi
\right) ^{2}}{4}-\frac{\pi ^{2}}{12},\text{ }x\in \left[ 0,2\pi \right] . 
\]
Thus the term $\delta \left( \phi \right) $ is given up to order $%
O(\varepsilon ^{3})$ by 
\begin{equation}
\delta \left( \phi \right) =\frac{8\varepsilon ^{3}}{3\pi }\left[ 2\frac{%
\sin ^{2}\left( \pi \eta /2\right) }{\left( \pi \eta /2\right) ^{2}}\cos
\left( 2\phi \right) -1+\frac{2}{\eta ^{2}}\left[ \eta +\frac{2}{\pi }\left|
\phi \right| -1\right] \Theta \left( \left| \phi \right| -\frac{\pi }{2}%
(1-\eta )\right) \right] .  \label{25a}
\end{equation}
where $\Theta \left( x\right) $ is the Heavside function, fulfilling $\Theta
\left( x\right) =1$ if $x>0$, $\Theta \left( x\right) =0$ if $x<0.$ (Of
course we shall take $\delta \left( \phi \right) \ =0$ if eq.$\left( \ref{20}%
\right) $ holds true.)

\section{\protect\smallskip Testable inequalities for the family of local
models}

A parameter giving a quantitative measure of the discrepancy between quantum
mechanics and the best model of the family defined by eqs.$\left( \ref{1}%
\right) $ and $\left( \ref{1a}\right) $ is, from eq.$\left( \ref{28}\right) ,
$%
\begin{equation}
\Delta ^{2}=\left\langle \delta \left( \phi \right) ^{2}\right\rangle \equiv 
\frac{1}{\pi }\int \delta \left( \phi \right) ^{2}d\phi =\frac{1}{2}%
\sum_{n=2}^{\infty }a_{n}^{2}\frac{\sin ^{4}\left( n\pi \eta /2\right) }{%
\left( n\pi \eta /2\right) ^{4}},  \label{35a}
\end{equation}
The summation in $n$ would be straightforward but lenghty and, as all terms
in the sum are positive, a lower bound is provided by the first few terms.
With just one term we get 
\begin{equation}
\Delta \geq \frac{\sqrt{2}\sin ^{3}\left( 2\varepsilon \right) }{3\left[
(\pi -2\varepsilon )\cos 2\varepsilon +\sin 2\varepsilon \right] }\frac{\sin
^{2}\left( \pi \eta \right) }{\pi ^{2}\eta ^{2}}\simeq \frac{8\sqrt{2}}{3\pi 
}\frac{\sin ^{2}\left( \pi \eta \right) }{\pi ^{2}\eta ^{2}}\varepsilon ^{3}.
\label{36}
\end{equation}
Actually it is easy to calculate $\Delta $ to order $O(\varepsilon ^{3})$
either performing the sum involved in eq.$\left( \ref{35a}\right) $ or
directly integrating the square of the expression $\left( \ref{25a}\right) $%
. I get 
\begin{equation}
\Delta =\frac{4}{3\pi }\sqrt{\frac{2}{3\eta }-\frac{1}{2}-\frac{\sin
^{4}\left( \pi \eta /2\right) }{\left( \pi \eta /2\right) ^{4}}}\left( V-%
\frac{\sin ^{2}\left( \pi \eta /2\right) }{\left( \pi \eta /2\right) ^{2}}%
\right) _{+}^{3/2}\equiv D(\eta ),  \label{36a}
\end{equation}
where I have taken $\varepsilon $ from eq.$\left( \ref{24a}\right) $ (as
before $\left( .\right) _{+}$ means putting zero if the quantity inside the
bracket is negative). The quantity $D(\eta )$ gives the deviation, as
defined in eq.$\left( \ref{25a}\right) ,$ between the best local model and
quantum mechanics. Thus in any experiment whose results are compatible with
the family of local models defined by eqs.$\left( \ref{1}\right) $ and $%
\left( \ref{1a}\right) ,$ the deviation will be greater than $D(\eta ).$
That deviation $\Delta _{\exp }$ may be obtained from the measurement of the
coincidence detection rates, $R_{12}(\phi ),$ at different angles, taking $%
\delta \left( \phi \right) $ to be the difference of that quantity
appropriately normalized minus the best cosinus fit of $p_{12}(\phi )$. Thus
we get 
\begin{equation}
\Delta _{\exp }=\left\langle \left[ \frac{R_{12}(\phi )}{\left\langle
R_{12}(\phi )\right\rangle }-1-V\cos 2\phi \right] ^{2}\right\rangle
^{1/2}\geq D(\eta ),  \label{30}
\end{equation}
whilst quantum mechanics predicts $\Delta _{\exp }=0$ . This inequality has
the virtue that it may be tested even at relatively low detection
efficiencies (provided that V is close enough to unity as is usual), so
removing the biggest obstacle in the standard tests of Bell\'{}s
inequalities. It replaces the previously proposed inequality (14) of Ref.%
\cite{Santos}, which is stronger but valid only for a family of LHV theories
more restricted than $\left( \ref{1}\right) $.

From the practical point of view the inequality $\left( \ref{30}\right) $
requires the measurement of the coincidence detection rate at different
values of the angle, $\phi ,$ between the polarizers. I may consider $n$
angles defined by 
\[
\phi _{j}=\frac{j}{n}\pi ,j=1,2...n, 
\]
and approximate the integral in the left hand side of $\left( \ref{30}%
\right) $ by a sum, so leading to the inequality 
\begin{equation}
\Delta _{\exp }=\left\{ \frac{1}{n}\sum_{j=1}^{n}\left[ \frac{R_{12}(\phi
_{j})}{\left\langle R_{12}\right\rangle }-1-V\cos 2\phi _{j}\right]
^{2}\right\} ^{1/2}\geq D(\eta ),  \label{30a}
\end{equation}
where 
\[
\left\langle R_{12}\right\rangle \equiv \frac{1}{n}\sum_{j=1}^{n}R_{12}(\phi
_{j}). 
\]

The quantity $V$ should be taken from the quantum prediction for the actual
experiment, but the inequality holds true for any value of $V$ provided we
use that same $V$ in the calculation of $D$($\eta ),$ eq.$\left( \ref{30}%
\right) .$ It is interesting to use the value of $V$ giving the minimum
value to $\Delta _{\exp }$ for the given empirical rates $\left\{
R_{12}(\phi _{j})\right\} ,$ that is 
\begin{equation}
V=2\frac{\sum_{j=1}^{n}R_{12}(\phi _{j})\cos 2\phi _{j}}{%
\sum_{j=1}^{n}R_{12}(\phi _{j})},  \label{30b}
\end{equation}
where I have taken into account the equalities 
\[
\sum_{j=1}^{n}\cos 2\phi _{j}=0,\sum_{j=1}^{n}\cos ^{2}2\phi _{j}=\frac{n}{2}%
.
\]
In this case I get 
\begin{equation}
\Delta _{\min }=\left\{ \frac{n\sum_{j=1}^{n}R_{12}^{2}(\phi _{j})-2\left(
\sum_{j=1}^{n}R_{12}(\phi _{j})\cos 2\phi _{j}\right) ^{2}}{\left(
\sum_{j=1}^{n}R_{12}(\phi _{j})\right) ^{2}}-1\right\} ^{1/2}\geq D(\eta ),
\label{30c}
\end{equation}
this inequality making possible empirical tests of the family of local
models defined by eqs.$\left( \ref{1}\right) $ and $\left( \ref{1a}\right) $
without any reference to quantum mechanics, whilst eq.$\left( \ref{30a}%
\right) $ allows tests of the local models versus quantum mechanics.

The test of the inequalities involve the measurement of $n/2+1$ rates if $n$
is even or $(n+1)/2$ rates if $n$ is odd, once we take into account the
symmetry $R_{12}(\phi _{j})=R_{12}(\pi -\phi _{j}).$ For even $n\geq 4,$ eq.$%
\left( \ref{30c}\right) $ may be related to the Fourier expansion eq.$\left( 
\ref{35a}\right) $ using the interpolation formula 
\begin{equation}
\frac{R_{12}(\phi _{j})}{\left\langle R_{12}\right\rangle }%
=1+\sum_{k=1}^{n/2}b_{k}\cos \left( 2k\phi _{j}\right) ,  \label{30E}
\end{equation}
whence the left hand side of $\left( \ref{30c}\right) $ may be interpreted
as 
\begin{equation}
\Delta _{\min }=\left\{ \frac{1}{2}\sum_{k=2}^{n/2-1}b_{k}^{2}+b_{n/2}^{2}%
\right\} ^{1/2}.  \label{30d}
\end{equation}

\smallskip The most simple non-trivial case corresponds to $n=4$, that is
the choice of the angles $0$, $\pi /4$ and $\pi /2.$ In this case eqs.$%
\left( \ref{30}\right) $ and $\left( \ref{30c}\right) $ lead to the
inequality

\begin{equation}
\Delta _{\min }=\frac{R_{12}\left( 0\right) +R_{12}\left( \frac{\pi }{2}%
\right) -2R_{12}\left( \frac{\pi }{4}\right) }{R_{12}\left( 0\right)
+R_{12}\left( \frac{\pi }{2}\right) +2R_{12}\left( \frac{\pi }{4}\right) }%
\gtrsim D(\eta ),  \label{25}
\end{equation}
Strictly speaking, from eq.$\left( \ref{30c}\right) $ we may only conclude
that the modulus of the left hand side fulfils the inequaltiy, but we may
take that side as positive because $a_{2}>0$ in the expansion $\left( \ref
{15a}\right) $. At a difference with $\left( \ref{30}\right) $ the
inequality $\left( \ref{25}\right) $ is not rigorous due to the
approximation made in going from eq.$\left( \ref{30}\right) $ to $\left( \ref
{30a}\right) ,$ the possible error being greater as the number of measured
rates is smaller.

In addition to the inequalities it is possible to derive some predictions of
the proposed local models for measurable quantities. For instance, defining 
\begin{equation}
V_{A}\equiv \frac{R_{12}\left( 0\right) -R_{12}\left( \frac{\pi }{2}\right) 
}{R_{12}\left( 0\right) +R_{12}\left( \frac{\pi }{2}\right) },V_{B}\equiv 
\sqrt{2}\frac{R_{12}\left( \frac{\pi }{8}\right) -R_{12}\left( \frac{3\pi }{8%
}\right) }{R_{12}\left( \frac{\pi }{8}\right) +R_{12}\left( \frac{3\pi }{8}%
\right) },  \label{40}
\end{equation}
their difference may be estimated. The former quantity is called the
visibility (or contrast) of the polarization correlation curve and the
latter is the quantity measured in the standard tests of Bell\'{}s
inequalities. According to quantum mechanics both quantities should be equal
to the parameter $V$, $\left( \ref{16}\right) $, but in our model there is a
difference. In fact, from eq.$\left( \ref{30E}\right) $ it follows that 
\[
V_{A}=\frac{b_{1}+b_{3}}{1+b_{2}+b_{4}}\simeq b_{1}+b_{3}-b_{2}-b_{4},V_{B}=%
\frac{b_{1}-b_{3}}{1-b_{4}}\simeq b_{1}-b_{3}+b_{4},
\]
where I have taken into account that, from a comparison of the empirical
quantities $b_{k}$ with the model quantities of eqs.$\left( \ref{28}\right) $
and $\left( \ref{24a}\right) ,$ we get 
\[
b_{4}\lesssim \left| b_{3}\right| \lesssim b_{2}<<1\simeq b_{1}\equiv V.
\]
Hence the family of LHV models here studied predicts (see eqs.$\left( \ref
{22}\right) $ and $\left( \ref{24a}\right) )$%
\begin{equation}
V_{B}-V_{A}\simeq b_{2}-2b_{3}+2b_{4}\simeq \frac{20\sqrt{2}}{3\pi }K\left(
V-\frac{\sin ^{2}\left( \pi \eta /2\right) }{\left( \pi \eta /2\right) ^{2}}%
\right) _{+}^{3/2},  \label{41d}
\end{equation}
K being a number of order unity. In some experiments it is reported, instead
of $V_{A}$, the value of $V$, which is obtained from a fit to the empirical
coincidence rates $R_{12}(\phi _{j}).$ In this case the model predicts 
\begin{equation}
V_{B}-V\simeq -b_{3}\simeq \frac{4\sqrt{2}}{3\pi }K\left( V-\frac{\sin
^{2}\left( \pi \eta /2\right) }{\left( \pi \eta /2\right) ^{2}}\right)
_{+}^{3/2}.  \label{41c}
\end{equation}
I stress that these estimates are correct only for the ``best'' local model
of the family. Thus their value is only indicative of the order of magnitude
to be expected in actual experiments if they are compatible with the said
family of local models.

In recent optical tests of Bell\'{}s inequalities people use two-channel
polarizers and four coincidence rates are measured, p$_{++}\left( \phi
\right) ,$ p$_{+-}\left( \phi \right) $, p$_{-+}\left( \phi \right) ,$ p$%
_{--}\left( \phi \right) .$ Typically there is symmetry between the two
channels, at least approximate. In this case our model may be extended to
these experiments using two functions $P_{+}\left( \phi \right) $ and $%
P_{-}\left( \phi \right) =P_{+}\left( \phi +\pi /2\right) $ instead of only
one, $P\left( \phi \right) ,$ as in eq.$\left( \ref{1}\right) .$ Assuming
the form $\left( \ref{21}\right) $ for $P_{+}\left( \phi \right) $ we get
the prediction 
\begin{equation}
R_{+-}\left( \phi +\pi /2\right) =R_{-+}\left( \phi +\pi /2\right)
=R_{++}\left( \phi \right) =R_{--}\left( \phi \right) ,  \label{41a}
\end{equation}
the latter given by eq.$\left( \ref{23}\right) $ (putting $\varepsilon =0$
if the inequality $\left( \ref{20}\right) $ holds true). The quantity
reported in the experiments is the correlation, defined by 
\begin{equation}
E\left( \phi \right) =\frac{R_{++}\left( \phi \right) +R_{--}\left( \phi
\right) -R_{+-}\left( \phi \right) -R_{-+}\left( \phi \right) }{R_{++}\left(
\phi \right) +R_{--}\left( \phi \right) +R_{+-}\left( \phi \right)
+R_{-+}\left( \phi \right) },  \label{42}
\end{equation}
for which our model predicts 
\begin{equation}
E\left( \phi \right) =V\cos (2\phi )-\delta \left( \pi /2+\phi \right) ,
\label{42c}
\end{equation}
where $\delta \left( \phi \right) $ is given by eq.$\left( \ref{25a}\right)
. $ The parameter most accurately measured in the experiments is 
\begin{equation}
S=\left| 3E\left( \pi /8\right) -E\left( 3\pi /8\right) \right| \equiv 2%
\sqrt{2}V_{B},  \label{42a}
\end{equation}
which is predicted to have the value $2\sqrt{2}$ according to quantum
mechanics in ideal experiments with 100\% detection efficiency, but S should
be less than 2 for any LHV model fulfilling the ``fair sampling hypothesis''
(see the Introduction section.)

From eq.$\left( \ref{40}\right) $ it is easy to see that, in our LHV family
of models, the quantities $V_{B}$ and $V_{A}$ defined in $\left( \ref{40}%
\right) $ correspond, respectively, to $2\sqrt{2}$ times the quantity S of
eq.$\left( \ref{42}\right) $ and to 
\begin{equation}
V_{A}=\frac{1}{2}\left( E(0)-E(\pi /2)\right) .  \label{42b}
\end{equation}
This allows testing the model prediction eq.$\left( \ref{41d}\right) $ in
experiments involving two-channel polarizers from measurements of the
correlation $E(\phi ),$ eq.$\left( \ref{42}\right) ,$ at the four angles $0,$
$\pi /8,3\pi /8$ and $\pi /2$. Also the prediction eq.$\left( \ref{41c}%
\right) $ may be tested getting $V_{B}$ from $\left( \ref{42a}\right) $ and $%
V$ from a fit of E($\phi )$ as shown in eq.$\left( \ref{42c}\right) .$

\section{From ideal to real experiments}

In the previous sections I have analyzed rather idealized experiments. In
actual polarization correlation experiments involving parametric down
converted photons the results obtained are not so simple as assumed in eqs.$%
\left( \ref{02}\right) $ and $\left( \ref{16}\right) .$ Indeed it is
frequent that the coincidence detection rate is not rotationally invariant
in the sense that $R_{12}\left( \phi _{1},\phi _{2}\right) $ does depends on
the angles $\phi _{1}$ and $\phi _{2}$ separately rather than on the
difference $\phi =$ $\phi _{1}-\phi _{2}$ only. In addition, the single
rates may depend on the positions of the polarizers, that is $R_{j}=R_{j}$($%
\phi _{j}),$ and the efficiencies, $\eta _{1}$ and $\eta _{2},$ of the
detectors may not be the same. Finally not all photons may arrive in pairs
at the detectors (ether because there is some single photon production or
because a fraction of the photons do not enter the apertures.) The relevance
of these nonidealities will be analyzed in the present section.

The quantities which may be measured in typical experiments involving single
channel polarizers are: Two single rates, $R_{1}$ and $R_{2},$ the rate $%
R_{12}$ of counts in the second detector conditional to counts in the first
detector (within an appropriate time window) and the quantum efficiencies of
the two detectors, $\eta _{1},\eta _{2}$ (these measured in auxiliary
experiments). In order to apply the analysis of Bell\cite{Bell} to the
experiments we should define single and coincidence probabilities$.$ But for
the test of the family of LHV theories studied in this paper (defined by eqs.%
$\left( \ref{1}\right) $ and $\left( \ref{1a}\right) )$ we do not need the
single probabilies and the single rates, $R_{j}$, are not used. In order to
obtain the coincidence probability, $p_{12},$ I introduce the rate, $R_{0}$,
of pair production in the source, whence I may get the $p_{12}$ using Bayes
rule of probability theory as follows 
\begin{equation}
p_{12}=p_{1}\times p_{2/1}=\frac{R_{1}}{R_{0}}\times \frac{R_{12}}{R_{1}}=%
\frac{R_{12}}{R_{0}}.  \label{44}
\end{equation}
The adequacy of this definition of $p_{12}$ requires that all photons
leaving the source are produced in pairs, with no more than one pair within
each time-window. That is, eq.$\left( \ref{44}\right) $ amounts at
neglecting single photon production, dark counts and the possible production
of several photon pairs within one time-window. I shall assume that all
these difficulties are avoided by a subtraction of ``accidental
coincidences'', which is a standard practice in polarization correlation
experiments. (I stress that this is so because we are attemting to test a
restricted family of LHV models, given by eqs.$\left( \ref{06}\right) ,$ but
would not be valid if we attempted to refute the whole family of LHV theories%
\cite{SantosFP}.) Actually the pair production rate $R_{0}$ cannot be
measured in the experiments but this is not a real difficulty. Indeed the
inequalities fulfilled by the LHV family of models proposed in this paper
involve only the dependence of the coincidence probability on the positions
of the polarizers, determined by the angles $\phi _{1}$ and $\phi _{2}.$ In
the experiments this dependence is usually studied by measurements with a
polarizer at a fixed position, say $\phi _{2}$ fixed, an the other
polarizer\'{}s position varied. Typically the measurements are made at two
different positions of the second polarizer, say $\phi _{2}=0$ and $\phi
_{2}=\pi /8.$ An appropriate interpolation leads to the following
parametrization 
\begin{equation}
p_{12}\varpropto R_{12}\varpropto 1+\left[ V_{m}+\left( V_{M}-V_{m}\right)
\cos 4\phi _{2}\right] \cos 2\phi ,  \label{45b}
\end{equation}
where $V_{M}(V_{m})$ is the m\'{a}ximum (minimum) visibility of the
detection coincidence curve when the angle $\phi _{+}$ is varied.

A local hidden-variables model able to reproduce eq.$\left( \ref{45b}\right) 
$ for not too high detection efficiencies is given by the functions $\rho $
and $P_{j}$ (see eq.$\left( \ref{06}\right) $) 
\begin{eqnarray}
\rho \left( \chi _{1},\chi _{2}\right)  &=&\frac{1}{\pi ^{2}}\left\{
1+\left[ W_{m}+\left( W_{M}-W_{m}\right) \cos 4\chi _{_{2}}\right] \cos
2\chi \right\} ,  \nonumber \\
P_{j}\left( \chi _{j},\phi _{j}\right)  &=&\beta _{j}\Theta \left( \frac{\pi
\eta _{j}}{4\beta _{j}}-\left| \chi _{j}-\phi _{j}\right| \right) ,\;j=1,2,
\label{45a}
\end{eqnarray}
with any $\beta _{j}\in \left[ \eta _{j},1\right] $ and appropriate choices
for $W_{M}$ and $W_{m}\leq $ $W_{M}.$ Here $\chi =$ $\chi _{1}-\chi _{2}$
and $\Theta \left( .\right) $ is the Heavside function $\Theta \left(
x\right) =1$ if $x\geq 0,$ zero otherwise. However if the detection
efficiencies are high enough and the value of $V_{M}$ in $\left( \ref{45b}%
\right) $ is close enough to unity, then it is not possible to reproduce the
quantum predictions. In fact, the conditions $\left( \ref{2}\right) $ imply $%
W_{M}\leq 1$ ( in addition to $\beta _{j}\leq 1)$ which give after some
algebra the constraint 
\begin{equation}
V_{M}\leq \frac{1}{3}\left( V_{M}+V_{m}\right) \left(
s_{1}^{2}+s_{2}^{2}\right) +\frac{4}{\pi ^{2}}\frac{s_{1}s_{2}}{\eta
_{1}\eta _{2}}\left[ 1-\frac{2}{3}\left( s_{1}^{2}+s_{2}^{2}\right) \right]
,\;s_{j}\equiv \sin \left( \frac{\pi \eta _{j}}{2}\right) .  \label{46}
\end{equation}
If $\eta _{j}<<1$ and both V$_{M}$ and V$_{m}$ are close to unity, as is
usually the case in typical experiments (where $V_{m}\gtrsim 0.96,\eta
_{j}\lesssim 0.3)$, the inequality becomes (compare with $\left( \ref{20}%
\right) )$%
\begin{equation}
V_{M}\lesssim 1-\frac{\pi ^{2}}{24}\left( \eta _{1}^{2}+\eta _{2}^{2}\right)
.  \label{46a}
\end{equation}
I stress that the quantities $\eta _{j}$ refer here to the efficiencies of
the photon detectors and not to the overall detection efficiency (which may
be substancially smaller than $\eta _{j}$ due to several losses). Agreement
between the model and quantum predictions is possible only if eq.$\left( \ref
{46}\right) $ (or eq.$\left( \ref{46a}\right) )$ is fulfilled. If this is
not the case, the model predicts inequalities which are violated by quantum
mechanics. In the following I study this case, which allows empirical
discrimination of the family of local models here proposed versus quantum
mechanics.

From the results of section 3 we see that the best choice for the functions $%
P_{j}$ is provided by eq.$\left( \ref{21}\right) $ (or, what is the same, eq.%
$\left( \ref{45a}\right) $ with $\beta _{j}=1.)$ ``Best'' is defined in the
sense that the disagreement with the quantum predicion is a minimum, the
disagreement being measured by an appropriate generalization of eq.$\left( 
\ref{7}\right) $. In order to get the best function $\rho $ we may assume
that the empirical results to be compared with the LHV model prediction are
obtained with a fixed $\phi _{2}$ $($see eq.$\left( \ref{45b}\right) ).$ In
this case all inequalities of section 4 are valid with the effective
visibility, $V_{eff}$, of the correlation curve substituted for $V$, where
we define 
\[
V_{eff}\equiv \left[ V_{m}+\left( V_{M}-V_{m}\right) \cos 4\phi _{2}\right] .
\]
It is easy to see that the stringent inequalities are obtained when $%
V_{eff}\equiv V_{M},$ corresponding to $\phi _{2}=0.$ Thus the inequalities
derived in the previous section apply, with $V_{M}$ substituted for $V$,
including the extension to experiments using two-channel polarizers and four
detectors (a trivial change is required if the efficiencies of the detectors
are not the same).

\section{Discussion}

Empirical tests of the models defined by eqs.$\left( \ref{1}\right) $ and $%
\left( \ref{1a}\right) $ are possible by means of the inequalities $\left( 
\ref{20}\right) $ and $\left( \ref{30a}\right) $, any experiment where one
of these inequalities holds true being compatible with a local hidden
variables model of the family. The tests are not difficult because
visibilities about 97\% and detection efficiencies of the order of 20\%
would be enough and these conditions have already been achieved in performed
experiments. For instance the experiment by Kurtsiefer et al.\cite{Kurt}
reports $V_{M}=0.982\pm 0.001$ and $V_{m}=0.970\pm 0.001$ after an
appropriate subtraction of accidentals. Any of these values, combined with
the efficiency $\eta =0.214,$ leads to a violation of the inequality $\left( 
\ref{20}\right) $, thus making the test possible using the inequalities $%
\left( \ref{30a}\right) $ or $\left( \ref{25}\right) .$ The quantities
reported in the published paper do not allow a test of the inequalities but
the experiment clearly shows that the empirical tests may be easily
performed with present technology.

The discrimination between the family of local models and quantum mechanics
is substantially more difficult than just to test the models. In fact,
calculating the predictions of quantum mechanics in actual (non-ideal)
experiments is far from trivial. For instance, in the ideal case the quantum
prediction is given by eq.$\left( \ref{16}\right) $ with both the detection
efficiency and the visibility of the coincidence curve equal to unity, i. e. 
$V=\eta =1.$ With these values both inequalities $\left( \ref{20}\right) $
and $\left( \ref{30a}\right) $ are violated, which excludes all local models
of the family here studied. This agrees with the well known fact that no
local hidden variables model is compatible with quantum mechanics for all
(ideal) experiments. Nevertheless the non-perfect behaviour of detectors
lowers their efficiency, $\eta ,$ making the results compatible with local
models in all experiments performed till now, also a well known fact\cite
{SantosFP}.

A restricted family of local models, like the one studied in this paper, may
be empirically refuted with a moderate value of the detection efficiency, $%
\eta ,$ combined with a relatively high value of the visibility, $V$,
provided that there is rotational symmetry, as assumed in section 3. However
in the fashionable parametric fluorescence experiments rotational invariance
does not hold true which makes easier the construction of local models
compatible with the experiments. In particular it is not obvious that
quantum mechanics, with all non-idealities taken into account, predicts a
detection counting rate of the form $\left( \ref{45b}\right) .$ If this is
not the case the fulfillement of the inequalities $\left( \ref{30a}\right) $
or $\left( \ref{25}\right) $ would not imply a violation of quantum
predictions.

In summary, the non-ideal behaviour of any actual experiment increases
substantially the range of parameters where quantum mechanics is compatible
with local hidden variables theories, thus making the empirical
discrimination rather difficult. It may even be the case that no actual, i.
e. non-ideal, experiment allows discriminating local hidden variables versus
quantum mechanics.

\end{document}